\documentclass[%
reprint,
 amsmath,amssymb,
 aip,
]{revtex4-2}

\usepackage{graphicx}% Include figure files
\usepackage{dcolumn}% Align table columns on decimal point
\usepackage{bm}% bold math
\usepackage{braket}% bold math
\usepackage{caption}
\usepackage{hyperref}
\usepackage{subcaption}
\usepackage{listings}
\usepackage{xcolor}
\usepackage{algorithmicx}

\newcommand{\bitleft}{\texttt{<<}}

\renewcommand{\and}{\texttt{\&}}

\begin{document}

\preprint{AIP/123-QED}

\title{\textbf{Approximate simulation of complex quantum circuits using sparse tensors} 
}% 

\author{Benjamin N. Miller}
 %\altaffiliation[Also at ]{Physics Department, XYZ University.}%Lines break automatically or can be forced with \\
\author{Peter K. Elgee}%

\author{Jason R. Pruitt}

\author{Kevin C. Cox}
 %\email{Contact author: Second.Author@institution.edu}
\affiliation{DEVCOM Army Research Laboratory\\
2800 Powder Mill Rd, Adelphi MD 20783}

\date{\today}% It is always \today, today,
             %  but any date may be explicitly specified

\begin{abstract}
%Simulation of quantum circuits using classical computers is a key avenue %for advancing our understanding and utilization of quantum systems.
The study of quantum circuit simulation using classical computers is a key research topic that helps define the boundary of verifiable quantum advantage, solve quantum many-body problems, and inform development of quantum hardware and software.  Tensor networks have become forefront mathematical tools for these tasks. Here we introduce a method to approximately simulate quantum circuits using sparsely-populated tensors. We describe a sparse tensor data structure that can represent quantum states with no underlying symmetry, and outline algorithms to efficiently contract and truncate these tensors.  We show that the data structure and contraction algorithm are efficient, leading to expected runtime scalings versus qubit number and circuit depth.  
Our results motivate future research in optimization of sparse tensor networks for quantum simulation. 
%Distribution level: A. Approved for public release; distribution unlimited. 
\end{abstract}

%\keywords{Suggested keywords}%Use showkeys class option if keyword
                              %display desired
\maketitle

%\tableofcontents
% \subsection{thesis sentences}
% -We introduce a method for approximate simulation of complex quantum circuits using sparse tensors.\\
% -Efficient tensor truncation gives truncation of paths\\
% -Complexity scales linearly with the path depth\\
% -Combines tensor network capabilities with approximate solutions\\
% -Decouples the basis size and gate depth with simulation complexity\\
% % -Feynman Path relative to MPS: show in math?\\
% -

%\section{Introduction}
Evaluating near-term applications of quantum computers is a vibrant field of current research. Quantum computers using trapped ions, superconducting qubits, atom arrays, and photonics are advancing toward near-term usefulness \cite{chen_benchmarking_2024, haghshenas_digital_2025-1, reichardt_fault-tolerant_2024, zhou_low-overhead_2025, muniz_high-fidelity_2025, psiquantum_team_manufacturable_2025, deng_gaussian_2023, aghaee_rad_scaling_2025}.
% xxcite? [google, quera, ionq, quantinuum]xx
Despite this progress, the engineering task of realizing fault-tolerant quantum computers is grand, with many outstanding challenges requiring large amounts of time and money.

On the other hand, classical simulators of quantum systems have made progress and are useful tools that are readily accessible now for studying quantum computational advantage and simulating important quantum many-body systems.  But strong complexity theoretical bounds exist for the difficulty of simulating quantum systems with classical computers in general.  Namely, P $\neq$ NP rules out the existence of a universal polynomial-time classical simulator. \cite{aaronson_complexity-theoretic_2017, boixo_characterizing_2018, hangleiter_computational_2023}  But many important problems are nonetheless tractable, and important progress has been made in many-body physics \cite{wu_variational_2024, pollmann_efficient_2016}, materials science \cite{perez-obiol_adiabatic_2022}, and, quantum chemistry \cite{shang_towards_2023}.  Methods for classically simulating quantum circuits have made impressive advances including traditional Schr{\"o}dinger evolution with parallel computing \cite{fatima_faster_2020, villalonga_establishing_2019, pan_simulation_2022} and Feynman-path style computation \cite{markov_quantum_2018,villalonga_establishing_2019}.  Tensor decomposition and renormalization techniques such as matrix product states \cite{zhou_what_2020}, density matrix renormalization group methods\cite{white_density_1992, ayral_density-matrix_2023}, and higher-dimensional tensor renormalization group methods \cite{levin_tensor_2007} allow one to make simplifying approximations based on underlying symmetries with significant impact in numerous research areas \cite{white_ab_1998, kurashige_entangled_2013,haghshenas_digital_2025}.
Most relevant to this work, tensor network-based quantum simulators are state-of-the art for simulating complex and random quantum circuits and have allowed orders-of-magnitude speedups in recent years that have paralleled the progress in quantum computing hardware. \cite{kourtis_fast_2019, gray_hyper-optimized_2021, pan_simulation_2022,  huang_efficient_2021, dubey_simulating_2025}.  
Here we introduce a simulation tool to further extend the capability of tensor network-based quantum simulators.  Our method, using truncated sparse tensors, may be used for tensor truncation and approximate simulation without the need for singular value decomposition, underlying symmetry, or low dimensionality.

In this paper, we first describe our method for approximate classical simulation of quantum circuits  using an approach we refer to as Truncated Sparse Tensor Simulation (TruSTS).  We describe an efficient data structure for TruSTS in the computational basis, and we derive and measure the resource cost of achieving a particular simulation fidelity $f$.  We show that simulation runtime scales linearly with the number of computed basis states $k$ and is roughly constant versus qubit number $N$ up to $N$=64. %pplication in a process we call 
This work points toward a future research avenue in sparse tensor network contraction and optimization, aiming to push the capability boundary for classical simulation of complex quantum systems.

%\section{Our Method of Simulation}

We begin by describing the TruSTS method.  We approximate the quantum state $\ket{\psi}$ of $N$ qubits using a sparse state $\ket{\phi}$, consisting of a linear combination of $k \leq 2^N$ basis states. We introduce a data-structure for storing $\ket{\phi}$, that we simply denote as $\phi$, to approximate the true quantum state $\ket{\psi}$ without allocating the full memory for the quantum state.  

In this work, we characterize the quality of the sparse state $\ket{\phi}$ using quantum fidelity $f = \lvert \braket{\psi | \phi} \rvert^2$.  If the quantum state $\ket{\psi}$ is inherently sparse, with many coefficients being zero or sufficiently small, the fidelity of the approximate state $f$ can remain large.  But it is also important to remember that while $f$ is a measure of $\ket{\phi}$'s overlap with $\ket{\psi}$'s mathematical wavefunction, it is not a measure of how well the system may emulate the corresponding quantum computer.  Just like a classical simulator, a quantum computer cannot perform state tomography (yielding $f$) in polynomial time, since that would require sampling all $2^N$ terms in $\ket{\psi}$.  Thus, although fidelity is a natural first choice for evaluating the performance of the quantum simulation in this initial work, investigations of sparse tensors as quantum state simulators using measurable quantities such as cross-entropy fidelity \cite{boixo_characterizing_2018} will be an important area for future research.

The sparse object $\phi$ consists of two k-element arrays ${\bf x}[x_i]$ and $\boldsymbol{\alpha}[\alpha_i]$.  ${\bf x}[x_i]$ is an array of integer basis-state coordinates $x_i$. The binary form of $x_i$ is equivalent to the bit string representation of the basis state $\ket{x_i}$.  We leverage this fact in subsequent sections by using bitwise operations on selected qubits in a manner similar to previous works \cite{fatima_faster_2020, jaques_leveraging_2021}. $\boldsymbol{\alpha}[\alpha_i]$ is an array that stores the quantum state's complex probability amplitudes $\alpha_i$ corresponding to the basis states $x_i$.  The sparse state array length $k$ is fixed before the simulation begins. %The process by which the array size is limited is detailed in the following section. 
Since $\ket{\phi}$ may contain less than $k$ nonzero elements, we introduce an integer pointer $n_{nz}$ to the last filled element of the arrays $\boldsymbol{\alpha}$ and $\bf{x}$%. to record the number of filled array elements.

% \subsection{Bitwise Contraction}\label{sec:contracton}

\begin{figure}
    \centering
    \includegraphics[width=1\linewidth]{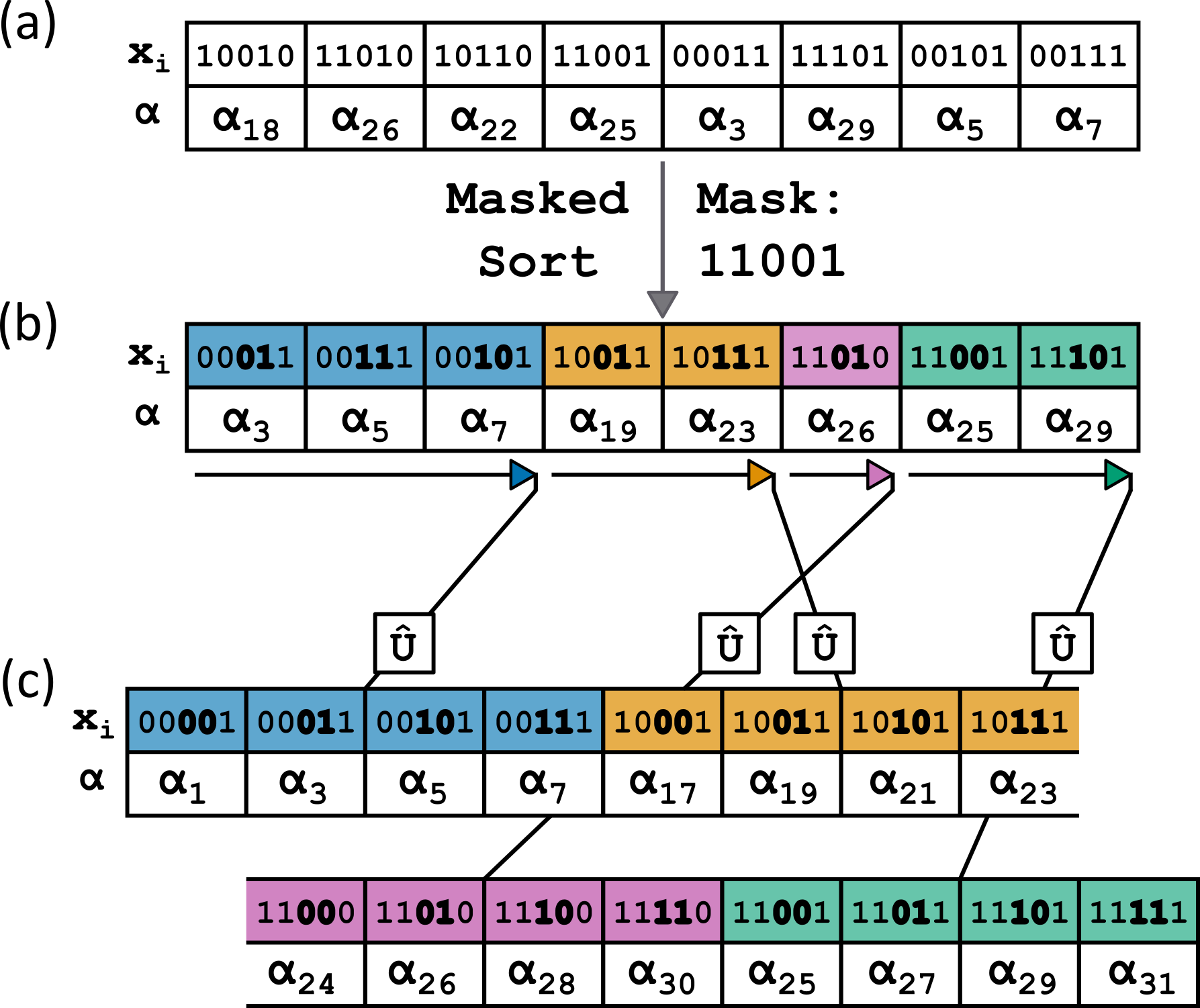}
    \caption{Diagram of the data structure and bitwise contraction algorithm for a sample state with $N = 5$ qubits, and $k = 8$.  The contraction is performed on lists of coordinates and data (shown in (a)) by sorting and grouping based on a bitmask (b).  After sorting, the gate may be applied independently to each group of coordinates (c). }
    \label{fig:contraction}
\end{figure}

Simulating a quantum circuit requires us to efficiently contract $\ket{\phi}$ with a tensor $\hat{U}$ that represents a quantum gate.  The data structure and tensor contraction method are displayed in Fig.~\ref{fig:contraction}.  Here we describe a two-qubit gate, but the algorithm generalizes to higher rank tensors in a straightforward manner.

We first define a bit mask to select the bits that are uncoupled by the gate $\hat{U}$.  The unmasked/masked values are indicated by bold/regular typeface in Fig. \ref{fig:contraction}.  These values are used to sort $\phi$ into computationally independent sub-arrays, shown by color. This sorting operation allows the gate $U$ to be applied via independent 4x4 matrix contraction on each sub-array. In general, application of the gate $\hat{U}$ may increase the number of nonzero elements $n_{nz}$ of $\ket{\phi}$ by up to a factor of four.  Source code and more detailed explanation of this contraction algorithm are included as Supplemental Material.

 % \subsection{Truncation and batching}
\begin{figure}[h]
    \includegraphics[width = 3.375 in]{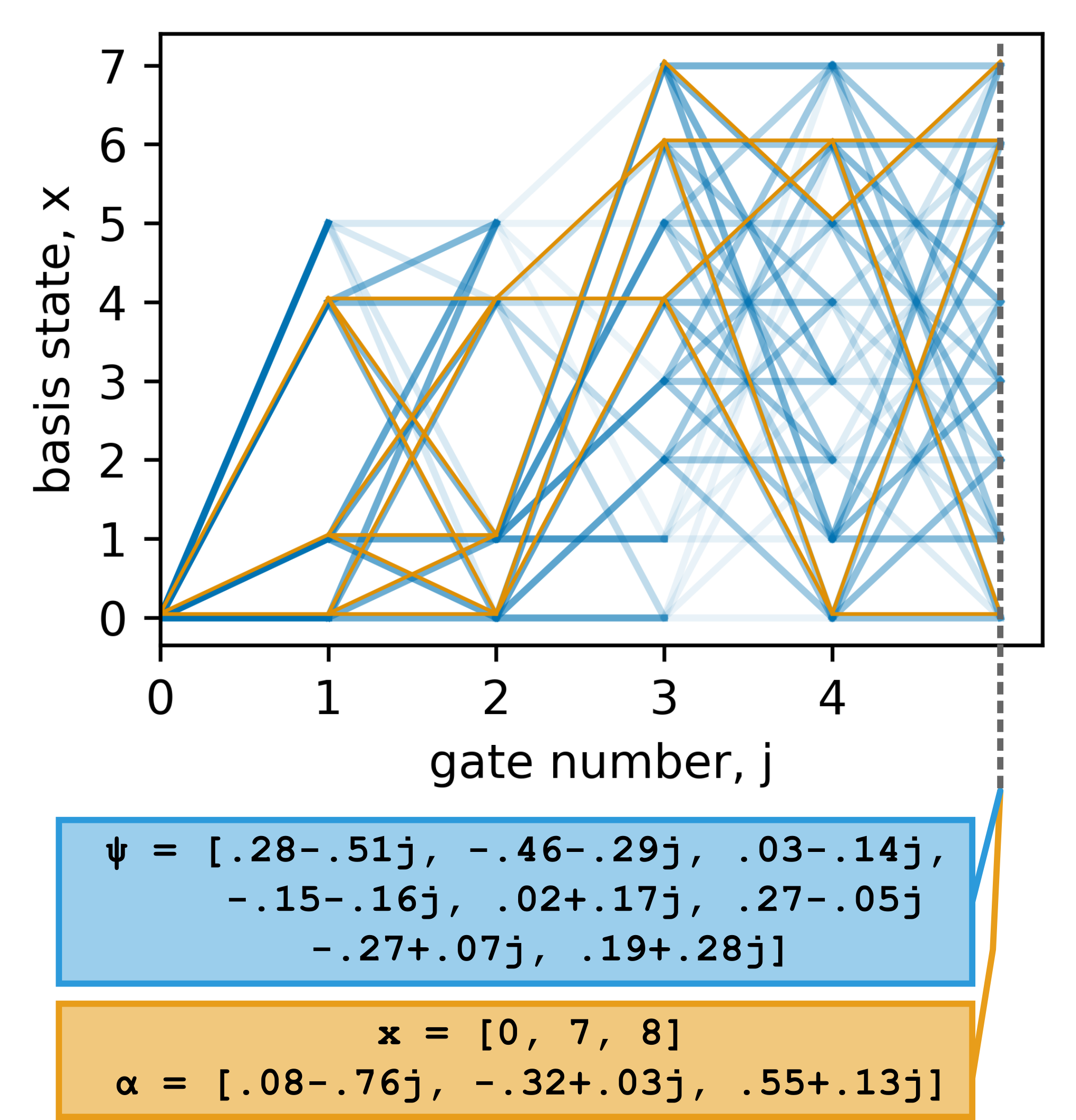}
    \caption{
    %Pictorial
    Representation of the TruSTS
    algorithm for $N=3$ qubits and $k = 2$, with sparse tensor terms represented as Feynman paths.
    The blue paths represent the terms in the exact circuit output shaded with an opacity corresponding to the probability amplitude, while terms kept in the sparse tensor truncation are shown in orange.  The exact and approximate final state are listed numerically below.}
    \label{fig:algorithm-graphic}
\end{figure}
Without truncation, subsequent application of gates leads to an exponential increase in $n_{nz}$.   Therefore, for the final step in TruSTS, after the contraction of $\hat{U}$, we remove % the smallest amplitude $\alpha_i$ entries in $\ket{\phi^j}$
$n_{nz}-k$ excess terms to recover the length-k sparse state when $n_{nz}>k$. In order to not overrun the $k$ terms of $\ket{\phi}$, we allocate an additional array of length of $4k$ to temporarily store the extra $n_{nz}-k$ elements before truncation.  In future work this fourfold memory overhead may be avoided by computing and truncating output terms in batches.  After truncation, we define the constant $\gamma<1$ that is needed to re-normalize $\ket{\phi}$ at the end of the algorithm.  $\gamma^2$ may be interpreted as the probability that is retained, or kept, through the multiple rounds of truncation.

In this work we primarily consider a truncation method which we refer to as ``top-$k$", where we remove the smallest $n_{nz}-k$ terms after each gate.  Additionally, we compare the performance of top-$k$ truncation to ``random-$k$" truncation, where $n_{nz}-k$ terms are removed at random.

An example evolution of a quantum state using the TruSTS algorithm is represented in Fig.~\ref{fig:algorithm-graphic} for $N=3$ qubits and length $k=2$. The initial state is chosen to be $\ket{\psi^0} = \ket{000}$. The circuit $\hat{U}_{\textrm{circuit}}$ is constructed with a product of five Haar-random 2-qubit unitary gates:
\begin{equation*}
    \hat{U}_{\textrm{circuit}} = \hat{U}^5\hat{U}^4\hat{U}^3\hat{U}^2\hat{U}^1
\end{equation*}
with gates applied sequentially over 2 randomly-chosen qubits. %using the tensor contraction algorithm outlined in the following section. 
The network of lines in Fig.~\ref{fig:algorithm-graphic} represents the quantum state evolution due to the application of each gate $\hat{U}^m$.  $\ket{\psi}$ is represented as blue lines and $\ket{\phi}$ is shown in orange.  The lines connect contributing basis state $x_i^{m}$ to the new basis states $x_j^{m+1}$ that result from the application of gate $\hat{U}^m$.  The line opacity from basis state $x^m_i$ to $x^{m+1}_j$ is given by the amplitude of each contracted term $\hat{U}_{ji}^m\cdot \alpha_{i}^m$.  The computational simplification afforded by the sparse state $\ket{\phi}$ is evident in the reduction of terms from $2^N$ to $k$.

This figure also emphasizes the differences and similarities between our sparse tensor contraction method and related Feynman path simulations \cite{markov_quantum_2018}.  In a Feynman path simulation of a circuit with $M$ gates, the final quantum state is calculated by summing over $4^M$ classical trajectories.  In contrast, amplitude paths that result in a particular basis state $x$ in TruSTS recombine.  That is, in each step,
\begin{equation}
    \alpha^{m+1}_j = \sum_i \hat{U}^m_{ji} \alpha^m_i.
\end{equation}
As a result, the number of terms $k$ in the sparse tensor representation will be less than or equal to the basis size $2^N$, rather than the potentially larger set of $4^M$ paths in a Feynman path approach \cite{markov_quantum_2018}.

% while a typical Feynman path approach involves sampling from a potentially larger set of $4^M$ paths \cite{markov_quantum_2018}.

We next derive and numerically show how TruSTS creates tradeoffs between memory, runtime, and state fidelity $f = \lvert \braket{\phi | \psi} \rvert ^2$.  We show that fidelity may be written in terms of the renormalization parameter $\gamma$.  This fact is useful since $\gamma$ %is normally computed during TruSTS and 
is an efficiently computed parameter even when TruSTS is applied to circuits which are intractable to compute exactly.  We perform this derivation in the Feynman path picture, considering all $4^M$ contributions to $\psi$, for simplicity.

We write the final state $\psi^M$  after $M$ quantum gates as a sum of $4^M$ Feynman paths:

\begin{equation}
    \ket{\psi^M}  = \sum_{i \in A}\sum_{j \in Q_i}a_j\ket{x_i}.
\end{equation}
where $A$ is the set of basis states, $a_j$ is the amplitude of path j, and $Q_i$ is the set of all Feynman paths that end in state $\ket{x_i}$, such that $Q = \bigcup_{i \in A} Q_i$ is the set of all $4^M$ Feynman paths.
%As we are not recombining paths, the set $Q$ is much larger than the dimension of the Hilbert space.

Next we consider truncating this state by only keeping some subset of paths $T \subset Q$.
% Due to the recombination of TruSTS, there is not a one-to-one correspondence between truncating paths, and truncating sparse tensor terms.\
The renormalization parameter after truncation is
\begin{equation}
    \gamma =\sqrt{ \sum_{i \in A}\left|\sum_{j \in T_i}a_j\right|^2},
\end{equation}
and the final truncated state $\ket{\phi_M}$ is
\begin{equation}
    \ket{\phi} = \frac{1}{\gamma}\sum_{i \in A}\sum_{j \in T_j}a_j \ket{x_i}.
\end{equation}

 Finally, we can compute the fidelity of this truncated state.
 \begin{equation}
     f = \left|\braket{\psi|\phi}\right|^2 = \frac{1}{\gamma^2}\left|\gamma^2 + \sum_{i \in A}\sum_{j \in Q_i\backslash T_i}a_j^*\sum_{k \in T_i}a_k\right|^2.
 \end{equation}

\begin{figure}
    % \centering
    \includegraphics[width=3.37in]{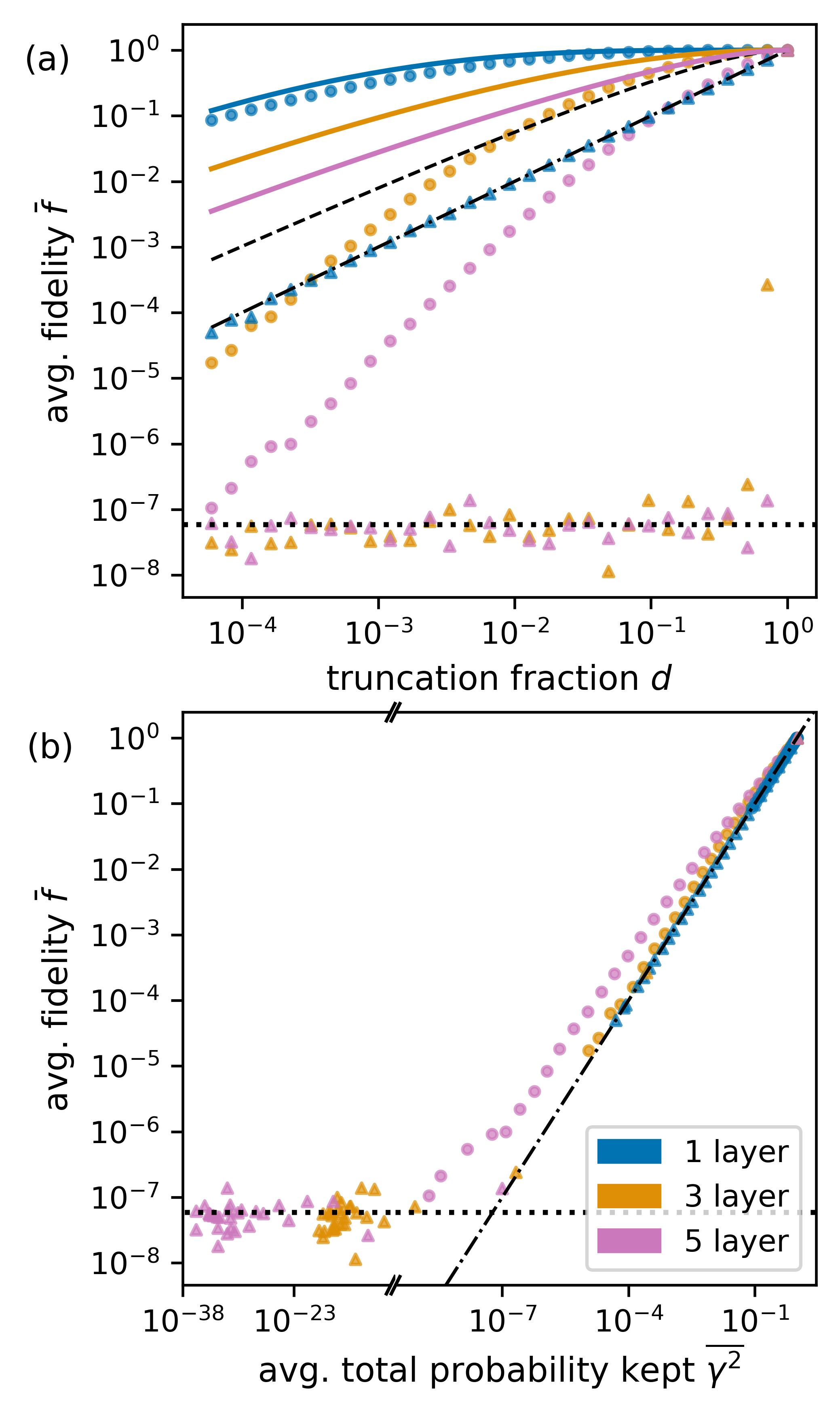}
    \caption{Final state fidelity vs the truncation fraction (a), and renormalization parameter $\gamma$ (b) for different layer numbers. In both (a) and (b), the square dotted gray horizontal line indicates the soft lower limit of fidelity $\bar{f}_\textrm{min}$ derived in Eq.~\ref{eq:fmin}, and circles represent top-$k$ truncation while triangles represent random-$k$ truncation. In (a), the solid lines represent the upper limit for the given layer number, and the dashed black line represents the upper limit for a circuit reaching a Porter-Thomas distribution as derived in Eq.~\ref{eq:tpresult}. In both (a) and (b), the diagonal dash-dotted black line has a slope of 1 and represents the expected fidelity derived in Eq.~\ref{eq:fidelityvsgamma}.}
    \label{fig:trunc_vs_fidelity}
\end{figure}

Ignoring the second term inside the absolute value, we recover $f = \gamma^2$, which provides a good guess at the expected fidelity of a given evaluation of our algorithm.
The second term encompasses all deviations from this simple relationship, and we can gain some insight into these deviations by examining its statistics.
This term involves the product of complex amplitudes that come from disjoint sets. 
Let us define
\begin{equation}
  \varepsilon = \sum_{i \in A}\sum_{j \in Q_i\backslash T_i}a_j^*\sum_{k \in T_i}a_k,  
\end{equation}
so that the fidelity becomes
\begin{equation}
  f = \frac{1}{\gamma^2}\left((\gamma^2 + \Re e(\varepsilon))^2+ \Im m(\varepsilon)^2\right)   
\end{equation}
For random quantum circuits, which we will consider in the next section, we expect amplitudes from the disjoint sets $T_i$ and $Q_i$ to be uncorrelated, and therefore the expectation value of $\varepsilon$ is zero.  If we assume that the real and imaginary parts of $\varepsilon$ have equal variances $\sigma_\varepsilon \ll \gamma^2$, on average the fidelity will be
\begin{equation}
    \bar{f} = \gamma^2 + \frac{\sigma_\varepsilon^2}{\gamma^2}> \gamma^2.
\label{eq:fidelityvsgamma}
\end{equation}
Thus, on average, we expect that the fidelity of our algorithm should be above the total probability kept after truncation.
This relationship is investigated empirically in Fig.~\ref{fig:trunc_vs_fidelity}(b).

We have calculated the fidelity as a function of $\gamma$ in the Feynman path picture. This result remains valid for TruSTS since we can draw a correspondence between truncating paths in the Feynman path framework and truncating sparse tensor terms in TruSTS.
Each truncation of a basis state $x^m_i$ in $\phi$ after application of gate $U^m$ corresponds to truncating all Feynman paths that pass through state $x_i$ after gate $U^m$.  Thus the calculation of $f$ with $\gamma$ is equivalent for both frameworks.

% \subsection{Fidelity limits}\label{sec:lower-limit}

While the above derivation gives us a relationship between $\gamma$ and the expected fidelity, and can help estimate the fidelity after running the algorithm and gaining access to the final approximate state, it does not predict the expected fidelity prior to running the algorithm.
Here, we provide additional limits on the fidelity based on algorithm parameters. %starting with a soft lower limit.

As an average worst-case, assume $k=1$ and we truncate to a single random basis state $\ket{j}$ with average fidelity (averaging over the choice $j$)

\begin{equation}
\bar{f} _\textrm{min}= 1/2^N
\label{eq:fmin}
\end{equation}
for any $N$-qubit circuit.
This provides the average fidelity we can expect when we truncate a significant portion of the possible states, or when the number of gates is large enough that the final state from the algorithm is fully uncorrelated with the true state. Achieving $\bar{f} _\textrm{min}$ is equivalent to making a random guess of the final state.

We also find an upper limit for the fidelity by truncating down to $k$ terms only after finding the true final state $\ket{\psi^M}$ rather than truncating mid-circuit.
This represents the best achievable fidelity of any state truncated to $k$ terms, and comparing to this value for a given $k$ provides a potentially useful figure of merit to benchmark the efficiency of TruSTS-like approaches in both this and future work.

We derive an expression for this fidelity in random circuits where the output distribution of probabilities for various bitstrings follows a Porter-Thomas distribution~\cite{porter_fluctuations_1956}, although this may be generalized to other circuits for which the output distribution is known. 
Consider a true final state $\ket{\psi} = \sum_i a_i \ket{x_i}$, where $p_i = |a_i|^2$ is the probability of measuring the final state in $\ket{x_i}$.
The probability density for measuring bit string $x_i$ with probability $p_i$ is
\begin{equation}
    Pr(p_i) = 2^Ne^{-2^N p_i}.
\end{equation}
Suppose we truncate $\ket{\psi}$ to a sparse state $\ket{\phi}$ containing the $k$ terms with the largest measurement probabilities. This represents a fraction $d = k/2^N$ of the total terms in $\ket{\psi}$, which we call the truncation fraction.
We can find the minimum probability of these $k$ terms by integrating the Porter-Thomas distribution
\begin{equation}
    d = \int_{p_{min}}^1 2^Ne^{-2^Np}dp
\end{equation}
yielding
\begin{equation}
    p_{min} \approx -\frac{\ln(d)}{2^N}.
\end{equation}

We now calculate the fidelity of the truncated state by summing the probabilities according to the expected probability distribution 
\begin{equation}
    f_\textrm{max} = \int_{p_{min}}^1pPr(p)dp \approx d(1 - \ln(d)).
\label{eq:tpresult}
\end{equation}
This represents the upper bound on the fidelity for any truncated state, and in general our algorithm's fidelity will fall below this value as we are truncating multiple times throughout the circuit. Reaching the fidelity $f_{max}$ implies that $\ket{\phi}$ contains the most probable bitstrings.  Therefore, while $f_{max}$ is potentially low, the truncated state may still be a high-quality simulator of the quantum circuit by returning the highest probability output states.%}As discussed in the introduction, this fidelity, while potentially low, represents a high-quality simulation of the corresponding quantum circuit since it represents a situation where the highest probability output states are returned.  Thus, a reasonable figure of merit for a quantum simulator may be the ratio of the achieved fidelity vs $f_{max}$.}

%\section{results}\label{sec:results}

% \subsection{Circuit Design}

\begin{figure}
    \centering
    \includegraphics[width = 3.375 in]{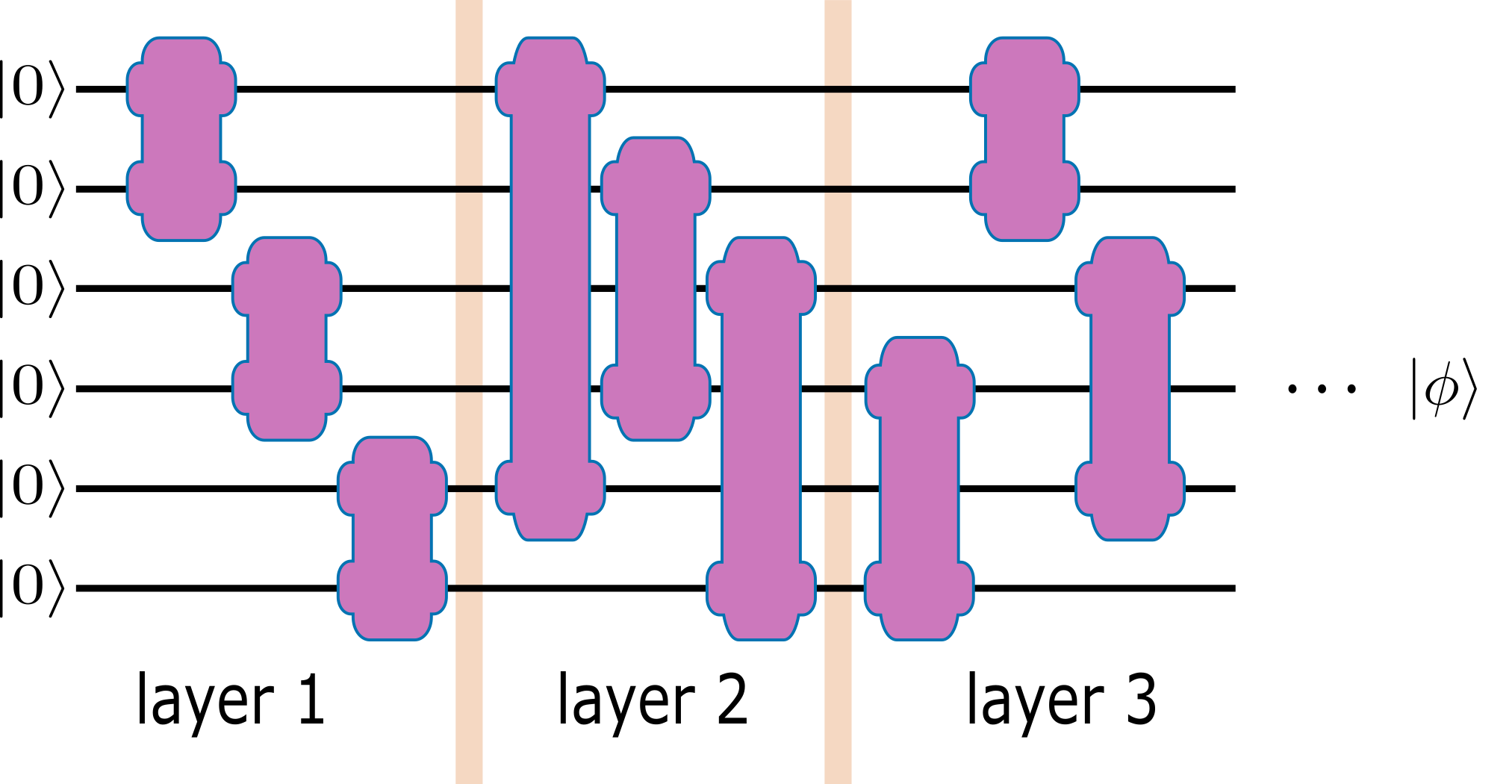}
    \caption{Example of the fully coupled random architecture
    %design, displayed
    with 3 
    layers shown.
    %with 3 2-qubit rotation gates in each. In general, this tiling pattern is repeated a number of times specified at runtime.
    }
    \label{fig:circuit_design}
\end{figure}

We benchmark the TruSTS approach on any-to-any random circuits consisting entirely of two-qubit unitary operators sampled from a Haar-uniform distribution. 
An example schematic with $N=6$ qubits of the random circuit architecture we use is shown in figure~\ref{fig:circuit_design}. 
Gates are arranged in a sequence of $L$ layers of $\frac{N}{2}$ gates (for even $N$) or $\frac{N-1}{2}$ gates (for odd $N$). 

Within each layer, gates are applied over randomly-paired qubits until all but at most one qubit have been coupled to. 
We choose this architecture as an example of a challenging problem class that is difficult to simulate using methods that exploit symmetry or low dimensionality such as matrix product states\cite{perez-garcia_matrix_2007}.

% Maybe include n-th layer??
% Label gates R_{q1, q2}?
% Spacing still not amazing - will fix

% \subsection{Fidelity}\label{sec:fidelity}

We calculate the fidelity of the TruSTS approximate state made of $k$ nonzero terms, $f=\lvert \braket{\phi^M|\psi^M}\rvert^2$. To ensure exact solutions can be computed on a typical desktop computer, we run these tests on circuits with only $N=24$ qubits. We compute this fidelity with a range of values for $k$, with corresponding truncation fraction $d = \frac{k}{2^N}$. $f$ is plotted versus $d$ and squared renormalization factor $\gamma^2$ for several cases in Fig.~\ref{fig:trunc_vs_fidelity}.

First, we compute $f$ for several relatively shallow circuits with layers $L = $1, 3, and 5.  The fidelity that results from TruSTS are shown as colored markers, where each data point is the average over ten different circuits. Averaged quantities are denoted with a bar.  The fidelity that results from top-$k$ truncation is shown in circles, and the fidelity that results from random-$k$ truncation is shown in triangles in Fig.~\ref{fig:trunc_vs_fidelity} (a) and (b).

We also show the upper and lower fidelity limits imposed by TruSTS in Fig.~\ref{fig:trunc_vs_fidelity}(a).  The lower limit of TruSTS fidelity $\bar{f}_\textrm{min}$ is shown as a blue dotted line.  The upper fidelity limits imposed by TRuSTS, when $\phi$ is only truncated at the end of the circuit, $f_\textrm{max}$, are shown as solid lines with color corresponding to the number of layers.  For the small layer counts in these tests, the resulting output distribution is not given by the Porter-Thomas formula, and we compute these bounds numerically.  Note that for a single layer of gates, TruSTS with top-$k$ truncation (blue circles) is approximately the upper bound.  For a deep random circuit whose output probabilities are described by the Porter-Thomas distribution, the upper limit is shown as a diagonal black dashed line using Eq.~\ref{eq:tpresult}.

Fig.~\ref{fig:trunc_vs_fidelity}(b) plots the measured average fidelity versus squared renormalization fraction, or probability kept, $\gamma^2$.  The data are seen to agree with the limits derived in Eq.~\ref{eq:fidelityvsgamma} and Eq.~\ref{eq:fmin} with $\bar{f}$ lying just above the line given by $f = \gamma^2$ and approaching the lower bound of fidelity in Eq. \ref{eq:fmin}.  In summary, the data of Fig.~\ref{fig:trunc_vs_fidelity} show that the fidelity of the approximate state $\ket{\phi}$ is given approximately by the probability kept $\gamma^2$. Top-$k$ truncation increases fidelity by concentrating probability in the kept terms.

% \subsection{Runtime Scaling}

\begin{figure}
    \centering
    \includegraphics[width = 3.375 in]{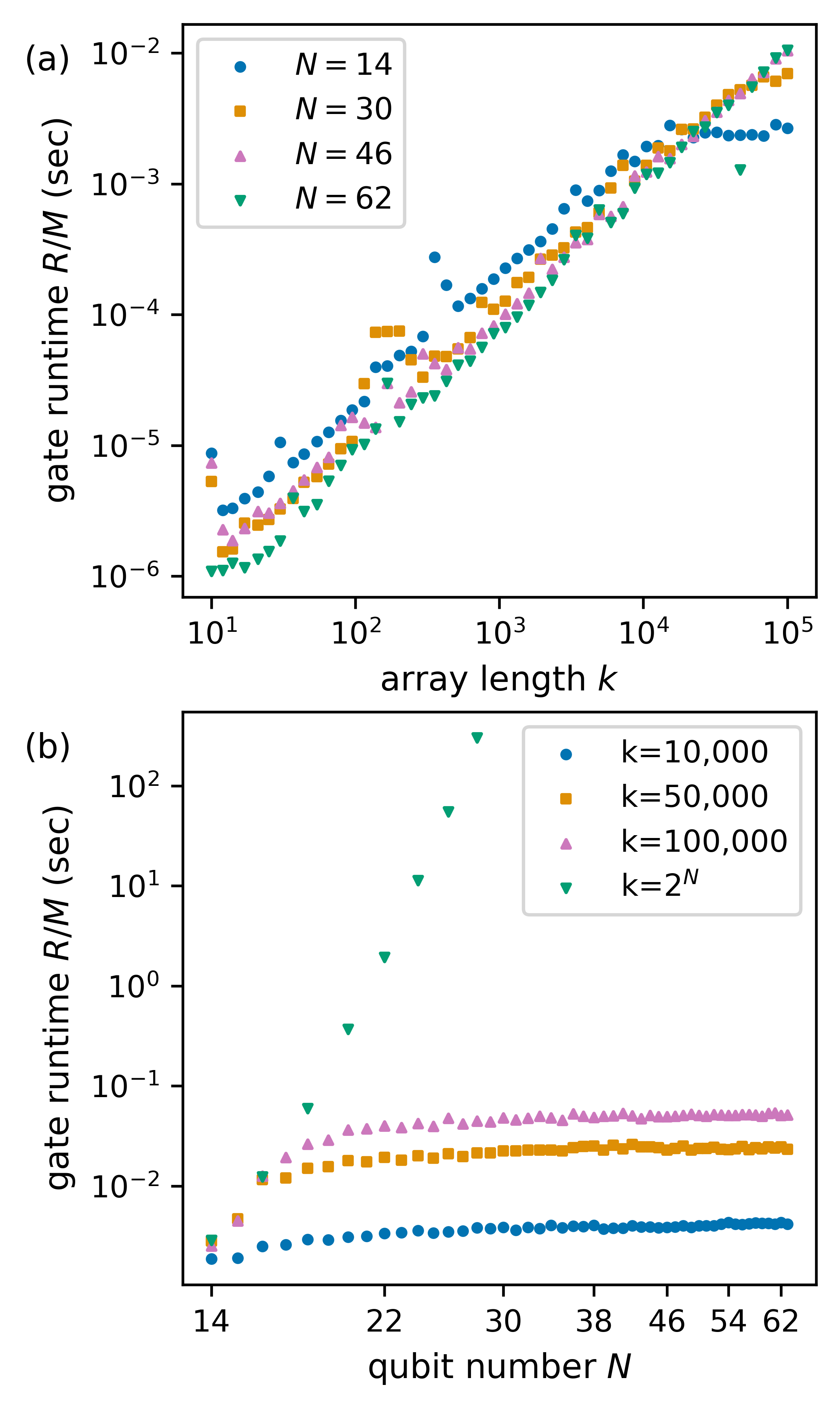}
    \caption{
    Runtime per gate vs $k$ (a), and qubit number (b).
    In (b) the green triangles represent the runtime per gate when running the exact circuit.
    }
    \label{fig:runtime_trust}
\end{figure}

Last, we experimentally quantify the runtime of the TruSTS algorithm.  
The results are shown in Fig. \ref{fig:runtime_trust}.  
The tests are performed using a WSL2 instance allocated 256 GB of DDR4 memory and an Intel(R) Xeon(R) w3-2345 CPU with 8 compute cores at a base frequency of 3.1 GHz. 
We measure the runtime $R$ as the difference of the ``wall-clock'' time before and after all steps of gate application and tensor truncation. 
Since the number of gates applied scales with the number of qubits for a constant number of layers $L$, we divide the runtime by the number of gates to get the run time per gate $R/M$. 

We test the algorithm's gate runtime scaling $R/M$ versus $k$ and $N$ by varying one and keeping the other constant, going from $N=14$ to $N=62$ in even steps to ensure that each qubit was operated on by a gate at each layer.  We run these tests on circuits with 5 layers each, and average the results over 10 such random circuits.%In both Fig.~\ref{fig:trunc_vs_fidelity}(a) and Fig.~\ref{fig:trunc_vs_fidelity}(b), each point corresponds to an average over 10 different random circuits with 5 layers. 
Fig.~\ref{fig:trunc_vs_fidelity}(a) demonstrates the expected result that $R/M$ scales linearly with $k$ for different qubit values $N$. 
Fig.~\ref{fig:trunc_vs_fidelity}(b) shows that runtime per gate $R/M$ is nearly constant with $N$ when truncating ($k << 2^N)$.  The non-constant trend in the low-$N$ case is a result of the number of nonzero values $n_{nz}$ being less than k.    When $k$ is increased to the maximum value $k = 2^N$ the runtime increases without observed bound.  
Together these plots demonstrate that our code implementation does not add additional runtime overhead beyond the fundamentally expected scalings, where the number of CPU operations performed is proportional to the number of terms to be multiplied. 
The sort operation that we rely on for TruSTS (see Fig.~\ref{fig:contraction}) is not observed to add significant overhead beyond the linear scaling for our data.

The data in Fig.~\ref{fig:runtime_trust} is run using TruSTS with $\textbf{x}$ stored as an array of 64-bit integers.  Running TruSTS with more than $N=64$ qubits requires a larger number of bits, and will lead to a runtime scaling that we expect to be linear with $N$.

%\section{Conclusion}
%-main path forward:  tensor network optimizations
%-many problems that are computationally easier than %random circuits

In this work, we have simulated random quantum circuits using straightforward state evolution and top-$k$ truncation.  We hope that our our bitwise-indexed sparse tensor contraction algorithm may find a variety of useful applications, extending the capabilities of available tensor packages such as ITensor or quimb \cite{gray_quimb_2018, fishman_itensor_2022}.   This first demonstration shows potentially useful tradeoffs between fidelity and runtime that may find immediate application in circuits where high fidelity is not required.  For particular quantum states that are inherently sparse, TruSTS is also attractive \cite{jaques_leveraging_2021}. Several additional details about TRuSTS, including a preliminary comparison between TruSTS and matrix product state simulation, are included as supplemental materials.  Rigorous comparison to alternate simulation methods is an important area for future research.

TruSTS may find an even more potent use case when combined with tensor network optimization strategies.  TruSTS naturally complements tensor network approaches which optimize over contraction order and therefore limit the number of deleterious truncations than the direct Schr\"odinger evolution implemented in this paper.  An interesting topic for future research will be tensor network optimization algorithms that explicitly consider the tradeoffs of sparsity and/or truncation.

Ultimately, our results suggest a number of promising avenues for future work into TruSTS, as well as other approaches with which TruSTS is complementary. We are optimistic that TruSTS can be a useful tool in exploring questions along the classical-quantum boundary.

\section*{Acknowledgments}
The authors acknowledge useful conversations with David Meyer, Paul Kunz, Josh Hill, Fredrik Fatemi, Cameron Taylor, and Kermit-James Leblanc.  

Research was sponsored by the Army Research Laboratory and was accomplished under Cooperative Agreement Number W911NF-26-2-A015.
The views and conclusions contained in this document are those of the authors and should not be interpreted as representing the official policies, either expressed or implied, of the Army Research Laboratory or the U.S. Government.
The U.S. Government is authorized to reproduce and distribute reprints for Government purposes notwithstanding any copyright notation herein.

\section*{Author Declarations}
\subsection*{Conflict of Interests}
The authors have no conflicts to disclose.

\subsection*{Author Contributions}
\textbf{Benjamin N. Miller:} Conceptualization (equal); Methodology (equal); Project administration (equal); Software (lead); Visualization (equal); Writing -- Original Draft (equal); Writing -- Review \& Editing (equal).  \textbf{Peter K. Elgee:} Conceptualization (equal); Methodology (equal); Project administration (equal); software (supporting); Visualization (supporting); Writing -- Original Draft (equal); Writing -- Review \& Editing (equal).  \textbf{Jason R. Pruitt:} Conceptualization (equal); Methodology (equal); Project administration (equal); Software (equal); Visualization (equal); Writing -- Original Draft (supporting); Writing -- Review \& Editing (equal).  \textbf{Kevin C. Cox:} Conceptualization (equal); Methodology (equal);  Project administration (equal); Software (supporting); Visualization (equal); Supervision (lead); Writing -- Original Draft (equal); Writing -- Review \& Editing (equal). 

\section*{Data Availability}
The data that support the findings of this article as well as the Python code that runs the calculations are available as supplemental material.

\bibliography{FeynmanPath.bib}% Produces the bibliography via BibTeX.
\section*{Supplemental Material}
\subsection{Bitwise Contraction Details}\label{app:contraction-algorithm}

Here we discuss the details of the general algorithm for contracting a sparse tensor with a dense gate in greater detail. We implement our entire circuit simulation code in python, using the numba library for fast, JIT-compiled (``Just In Time") code where there is meaningful benefit \cite{lam_numba_2015}, such as in the contraction itself. Assume that the sparse state $\ket{\phi}$ is represented using two arrays of fixed length $k$: a coordinate array of unsigned integers ${\bf x}[x_i]$ and an array of complex-valued coordinates $\boldsymbol{\alpha}[\alpha_i]$, such that $\alpha_i$ is the probability amplitude corresponding to basis state $x_i$. $x_i$ is the unsigned integer representation of a bitstring corresponding to a particular basis state in $\{\ket{0},\ket{1}\}^N$, which might be represented as an $N$-length boolean in a typical sparse tensor. The array length is fixed throughout the simulation to avoid slow memory reallocation. Since in general, $\ket{\phi}$ might have fewer than $k$ elements, we define the number of nonzero values $n_{nz}\leq k$ to point to the end of the nonzero terms of ${\bf x}$ and $\boldsymbol{\alpha}$. The 2-qubit gate $\hat{U}$ is a length-$16$ array $\mathbf{U}[U_j]$ of complex values.  We leverage bitwise representations of indices of $\mathbf{U}$ to compare with elements of ${\bf x}$.

The basic contraction algorithm is inspired by multiplication of a Compressed Sparse Row (CSR) matrix \cite{saad_iterative_2003}. It consists of sorting using $\bf x$ to group separate independent subspaces in the basis (``rows'' in CSR), then applying a gate $\hat{U}$ to each subspace. Sorting allows the entire gate to be applied with a single memory traversal, rather than searching for terms which might recombine. Assume a gate is applied over qubits $(g_1,g_2)$ with corresponding ``gate mask'' defined as $m_g = (1\bitleft g_1) ~\texttt{\&}~ (1\bitleft g_2)$, where $\bitleft$ is a bitwise shift left, and $\texttt{\&}$ is a bitwise \texttt{and} operation. For example, for a gate applied over qubits $1$ and $4$, $m_g=0010010=18$. For each $x_i$, we first define $x^f_i = x_i~\texttt{\&} ~ \mathord{\sim}m_g $ and $x^c_i=x_i~\texttt{\and}~ m_g$, where $\sim$ is the bitwise \texttt{not} operator. The sorting step is implemented using a quicksort algorithm which uses $x^f_i$ as the sort key and sorts both $\bf x$ and $\boldsymbol{\alpha}$ together. Note that each unique $x^c$ constitutes a distinct subspace in the basis that is independent under the action of $\hat{U}$, and all basis states $x_i$ which share a subspace are sequential within $\bf x$ after sorting. 

To apply the gate $\hat{U}$ to $\phi_i = (x_i, \alpha_i)$, we first separate $x_i$ into $x^f_i$ and $x^c_i$. Then, for each index $j_{c}$ in $\hat{U}$ whose binary representation starts (from the right) with $x^c_i$, the output array elements $\alpha_{out} = \alpha_i\cdot U[j_c]$. To compute each output coordinate $x_{out}$,
we replace $x^c_i$ with the last two digits of the binary representation of $j_c$.  
Terms with equivalent $x_{out}$ need to be combined.  This is simplified by the fact that the sort groups all terms with equivalent $x^f_i$ and searching $\bf x$ is not necessary for this step.

\subsection{Comparison With MPS}
Matrix Product States (MPS) are another common representation of quantum states useful for approximate simulation of quantum circuits  \cite{perez-garcia_matrix_2007}.  Here, we perform a preliminary comparison of fidelity and runtime between TruSTS and MPS.
 
We empirically demonstrate the resource tradeoffs for both the MPS approach and TruSTS and discuss fundamental differences between the outputs of the two methods. 

\begin{figure}
    \centering
    \includegraphics[width= 3.375 in ]{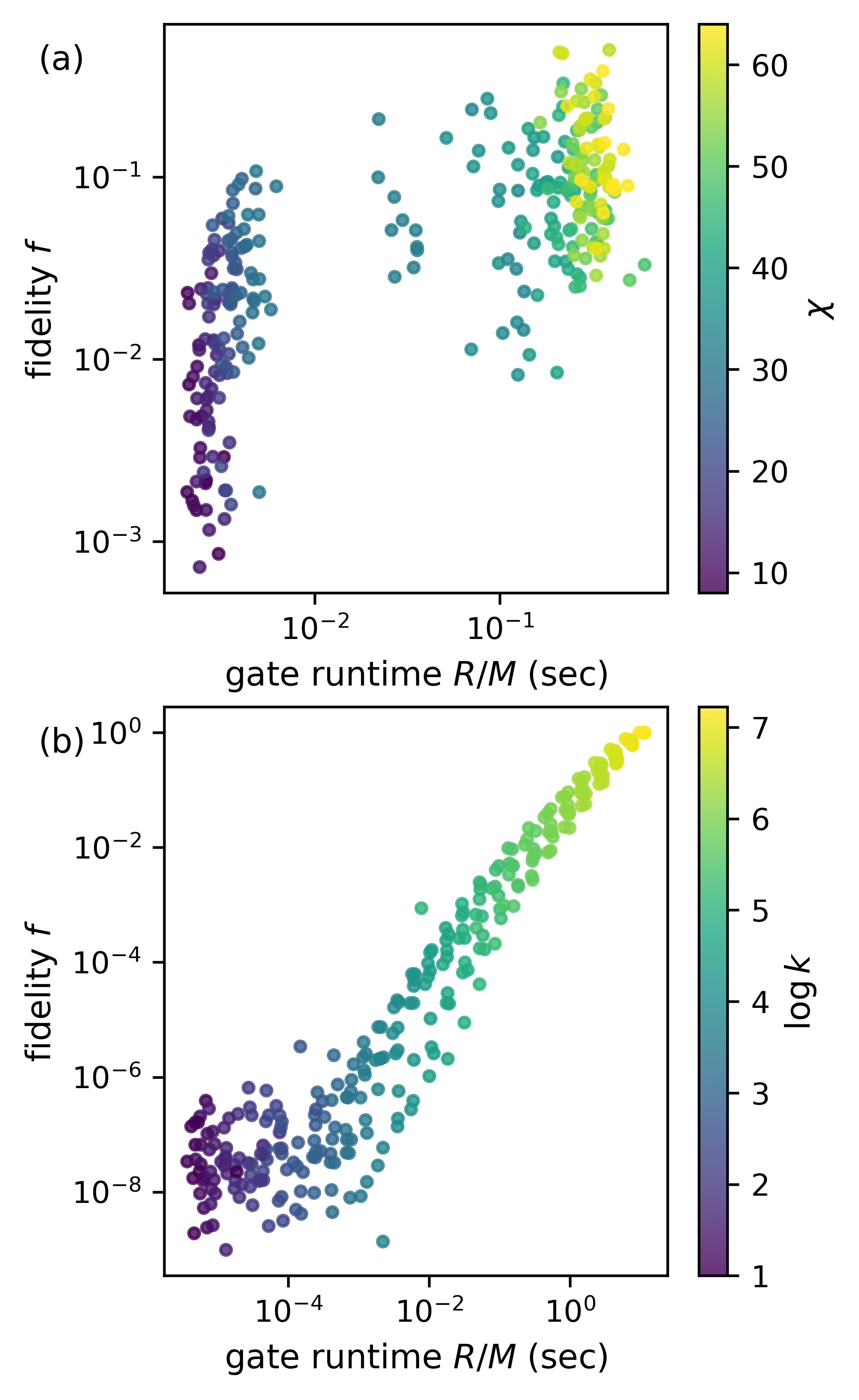}
    \caption{Fidelity $f$ vs runtime $R/M$ for (a) MPS (using the quimb package) and (b) TruSTS for a $N=24$ qubit circuit with $l = 5$ layers. In (a), the colorbar indicates the maximum bond dimension $\chi$, and in (b) the colorbar indicates the number of terms $k$ kept after truncation.}
    \label{fig:mps_compare}
\end{figure}

To run circuits using MPS, we use the CircuitPermMPS implementation built into the quimb library \cite{gray_quimb_2018}, using equivalently-generated random circuits to those used in our test of TruSTS. In contrast to the $k$ parameter in TruSTS, the tradeoff between performance and expressiveness of MPS is dictated by its maximum bond dimension $\chi$. For this comparison, we test the 5-layer case, increasing $\chi$ and tracking the subsequent changes in runtime and fidelity. Notably, MPS can express a many-body state of $N$ 2-level systems perfectly with $\chi=2^{\frac{N}{2}}$. For a given circuit, we first compute the output with an exact state $\ket{\psi}$ represented as a MPS with this $\chi=2^{\frac{N}{2}}$. We then run the same circuit with a restricted $\chi$ to get an approximate output state $\ket{\phi_{MPS}}$, and calculate the fidelity $f=\left|\braket{\phi_{MPS}|\psi}\right|^2$ by joining the tensor network representations of $\ket{\phi_{MPS}}$ and $\ket{\psi}$ and contracting. The fidelity data are plotted against gate runtime for MPS simulations and TruSTS in figure~\ref{fig:mps_compare}. The gate runtime R/M is defined as the difference in wall-clock time between the simulation start and end divided by the number of gates.

For $N=24$, $l=5$, Matrix Product States have a more-favorable tradeoff between runtime and fidelity compared with TruSTS over most of the data.  For the parameters chosen, the fidelity of the MPS remains above $10^{-2}$ for $\chi >20$.  For lower $\chi$, there is a rapid dropoff in fidelity.  The fidelity of TRuSTS decreases more quickly and steadily versus runtime, until it reaches the lower limit near $f \sim 10^{-8}$.

% \bibliography{FeynmanPath.bib}

\end{document}